\documentstyle[12pt]{article}

\newlength{\extraspace}
\newlength{\extraspaces}
\newcommand{\be}{\begin{equation}
\addtolength{\abovedisplayskip}{\extraspaces}
\addtolength{\belowdisplayskip}{\extraspaces}
\addtolength{\abovedisplayshortskip}{\extraspace}
\addtolength{\belowdisplayshortskip}{\extraspace}}
\newcommand{\ee}{\end{equation}}

\newcommand{\bq}{\begin{eqnarray}
\addtolength{\abovedisplayskip}{\extraspaces}
\addtolength{\belowdisplayskip}{\extraspaces}
\addtolength{\abovedisplayshortskip}{\extraspace}
\addtolength{\belowdisplayshortskip}{\extraspace}}
\newcommand{\eq}{\end{eqnarray}}

\newcommand{\newsection}[1]{
\vspace{15mm}
\pagebreak[3]
\addtocounter{section}{1}
\setcounter{equation}{0}
\setcounter{subsection}{0}
\setcounter{footnote}{0}
\begin{flushleft}
{\large\bf \thesection. #1}
\end{flushleft}
\nopagebreak
\medskip
\nopagebreak}

\begin{document}
\hbox{}
\nopagebreak
\vspace{-3cm}
\addtolength{\baselineskip}{.8mm}
\baselineskip=24pt
\begin{flushright}
{\sc LA-UR}-94-2727\\
{\sc PUPT}- 1492\\
hep-th@xxx/9408081 \\
 August  1994
\end{flushright}

\begin{center}
{\Large  A variational approach to the QCD wave functional:\\
 Dynamical mass generation and confinement.}\\
\vspace{0.1in}
{\large Ian I. Kogan}
\footnote{ On  leave of absence
from ITEP,
 B.Cheremyshkinskaya 25,  Moscow, 117259,     Russia.\\
 Address after August 15, 1994: Department of Theoretical
 Physics, 1 Keble Road, Oxford, OX1 3NP, UK}\\
{\it  Physics Department, Princeton  University \\
 Princeton, NJ 08544
 USA} \\
\vspace{0.1in}
{\large Alex Kovner}\footnote{ Address after September 15, 1994:
 Physics Department, University of Minnesota,
 116 Church Street S.E., Minneapolis, MN 55455, USA}\\
{\it  Theory Division, T-8,  LANL\\
 Los-Alamos, NM 87545,
 USA}

\vspace{0.1in}

{\sc  Abstract}
\end{center}

\noindent
We perform a variational calculation in the SU(N) Yang Mills
theory in 3+1 dimensions.  Our trial variational states
are explicitly gauge invariant, and reduce to simple Gaussian
states in the zero coupling limit. Our main result is that the
energy is minimized for the value of the variational parameter
away form the perturbative value. The best variational state
is therefore characterized by a dynamically generated  mass scale $M$.
This scale is related to the perturbative scale $\Lambda_{QCD}$
by the following relation: $\alpha_{QCD}(M)={\pi\over 4}{1\over N}$.
Taking the one loop QCD $\beta$- function and $\Lambda_{QCD}=150 Mev$
 we find (for N=3) the vacuum condensate
${\alpha\over \pi}<F^2>= 0.008 Gev^4$.
\vfill

\newpage
\newsection{Introduction.}

\renewcommand{\footnotesize}{\small}

Understanding of low energy phenomena in QCD, such as
confinement and chiral symmetry breaking
  or, in more general terms,
  the strong coupling problem  and  a ground state structure
  in an asymptotically free non-abelian gauge theory is, without
 doubt, one of  the main (if not the main)
  problems in  modern quantum field theory. In spite of years
  of attempts to answer this question we are still far from
  complete satisfaction, although a lot of interesting
 and promising ideas were suggested during the first $20$
 years of QCD \cite{20years}.

Considerable progress  has been made in this direction during
last years
using numerical approach of the lattice gauge theory \cite{lattice}.
The lattice gauge theory calculations are, however, still incomplete.
Apart from that, they sometimes leave behind an unpleasant aftertaste
(although this is of course, a very subjective matter)
that one obtains numerical results without gaining real
understanding of the  underlying physics.
To our minds, understanding of these issues in the framework of
an analytical approach
would be invaluable. An analytic method that is capable of solving
the low energy sector of QCD starting from first principles,
would also, presumably,
teach us a lot about other strongly interacting theories such
as technicolor.

Unfortunately, the arsenal of nonperturbative methods to
 tackle strongly interacting continuum
theories is very limited, to say the least.
Methods that perform very well in simple quantum mechanical problems,
are much more difficult to use in quantum field theory (QFT).
This is true,
for example, for
a variational approach. In quantum mechanics it is usually enough
to know a few simple qualitative features in order to set up a
variational ansatz which gives
pretty accurate results, not only for the energy of a ground
state, but also
for various other vacuum expectation values (VEV).
In QFT one immediatelly is faced with several difficult
problems when trying to apply this method, as discussed
insightfully by  Feynman \cite{vangerooge}.

First, there is the problem of calculability. That is, even if one
had a very good guess at the form of the vacuum wave functional (or,
for that matter,  even knew its exact form) one would still have to
evaluate expectation values of various operators in this state. In a
field theory in d spatial dimensions this involves performing a
d- dimensional path integral, a problem, in itself very complicated and,
in general not manageable. This problem is especially severe in
nonabelian gauge theories, where gauge invariance poses strong
restrictions on admissible trial wave functionals (WF). In this
case it becomes very difficult to find a set of WF's which are both,
gauge invariant and amenable to analytic calculation.

Another serious problem is the problem of "ultraviolet modes". This
means the following. In a variational calculation of the kind we have
in mind, one is mostly interested in the information about the low
momentum modes. However, the VEV of the energy (and all other
intensive quantities) is dominated entirely by contributions of
high momentum fluctuations, for a simple reason, that there are
infinitely more UV modes than modes with low momentum. Therefore,
even if one has a very good idea, how the WF at low momenta should
look like, if the UV part of the trial state is even slightly
incorrect the minimization of energy may lead to absurd results.
Due to the interaction between the high and low momentum modes,
the IR variational parameters will in general be driven to values
which minimize the interaction energy, and have nothing to do with
the dynamics of the low momentum modes themselves.

Even though over the years many attempts at variational calculations
in QCD have been made \cite{variational}, these two problems
invariably made their presence felt, and at this point one really
can not point to any succesful variational calculation in a
nonabelian gauge theory.
Our feeling is,
however, that these obstacles are not necessarily insurmountable,
and that this direction is still far from being
exhausted and deserves further development.

In this paper we present a variational calculation of the vacuum
WF in a pure SU(N) Yang Mills theory, which, at least partially,
is free from the problems mentioned earlier. We use wave functionals
that are explicitly gauge invariant.
The correct UV behaviour is built into our ansatz. In the case at
hand this can be done due to the assymptotic freedom of the models
considered. We are able to calculate VEVs of local operators in our
trial states in a reasonable approximation, combining the renormalization
group and the mean field techniques.

Our main result is that the energy is minimized at the value of the
variational parameter away from the perturbative vacuum state.
This leads to a dynamical generation of scale in the vacuum WF. The
value of the vacuum condensate ${\alpha\over\pi}<F_{\mu\nu}^{a}
F_{\mu\nu}^{a}>$ in this state
turns out to be equal to $0.008 Gev^4$ for $\Lambda_{QCD}=150 Mev$.
Even though this result should be taken only as an order of
magnitude estimate (due to approximations made), it is pleasing
to see a number so close to the phenomenologically known
$0.012 Gev^4$ \cite{sumrules} emerge from this simple calculation.

The paper is organized as follows. In Section 2 we set up the
variational calculation and discuss in some detail our variational
ansatz. In Section 3 we discuss the approximation scheme for
calculation of VEVs in the trial WF. In Section 4 the minimization
of energy and calculation of $<F^2>$ are performed. Some elements of the
calculation of the Wilson loop  and an area law is discussed in
 Section 5. Finally, Section 6
contains dicussion of our results and outlines directions for future work.

\newsection{The variational trial state and the gauge invariance.}

As mentioned in the previous section, an immediate question one is
faced with, when picking a possible variational state is calculability.
One should be able to calculate averages of local operators in this
state
\begin{equation}
<O>=\int D\phi \Psi^*[\phi]O\Psi[\phi]
\end{equation}
A calculation of this kind obviously, is tantamount to evaluation
of a Euclidean path integral with the square of the WF playing the
role of the partition function. One should therefore be able to
solve exactly a $d$-dimensional field theory with the action
\begin{equation}
S[\phi]=-{\rm log}\Psi^*[\phi]\Psi[\phi]
\end{equation}
Since in dimension $d>1$ the only theories one can solve exactly
are free field theories, the requirement of calculability almost
unavoidably restricts the possible form of the WF to a Gaussian
(or as it is sometimes called squeezed) state:
\begin{eqnarray}
\Psi[\phi]=\exp\left\{-{1\over 2}
\int d^{3}x d^{3}y \left[ \phi(x)-\zeta(x)\right]
G^{-1}(x,y)\left[\phi(y)-\zeta(y)\right]\right\}
\label{gaus}
\end{eqnarray}
with $\zeta(x)$ and $G(x,y)$  being c-number functions.
 The requirement of translational invariance usually gives
further restrictions: $\zeta(x)=\rm const$,~ $G(x,y)=G(x-y)$.

The restriction to Gaussian WF is of course a severe one. However,
one can still hope that in some cases the simple Gaussian form can
capture the most important nonperturbative characteristics of the
true vacuum. Indeed, the Gaussian variational approximation has
been used succesfully in self interacting scalar theories, where
it is known to be exact in the limit of large number of
fields. Perhaps the most celebrated use of these trial
states is the BCS calculation of the superconducting ground state
\cite{book}, where for most of the interesting quantities its
accuracy is of order 10-20\%.

The reason the approximation works well in these theories, is that
in both cases a single condensate dominates the nonperturbative
physics, and the Gaussian ansatz is wide enough to accomodate this
most important condensate.
 From this point of view, it would seem then, that it is perfectly
reasonable to
try a similar variational ansatz in the Yang - Mills theory. After all,
it is strongly suggested by the QCD sum rules \cite{sumrules} that the
pure glue sector is dominated by one nonperturbative condensate $<F^2>$.
 We also know, that the VEV of the field strength itself $<F>$ vanishes,
since it is not a gauge invariant operator. A state of the form
(\ref{gaus}) with $\zeta=0$ would indeed give zero classical fields,
but nonzero quadratic condensates.

There is, however, one obvious difficulty with this idea. It is very
easy to see, that in a nonabelian theory it is impossible to write down
a Gaussian WF which satisfies the constraint of gauge invariance.
The SU(N) gauge theory is described by a Hamiltonian
\begin{equation}
H= \int d^{3}x \left[{1\over 2}E^{a2}_i+{1\over 2}B^{a2}_i\right]
\label{ham}
\end{equation}
where
\begin{eqnarray}
E^a_i(x)&=&i{\delta\over \delta A^a_i(x)} \nonumber \\
B^a_i(x)&=&{1\over 2}\epsilon_{ijk}
\{\partial_jA_k^a(x)-\partial_kA^a_j(x)+gf^{abc}A_j^b(x)A_k^c(x)\}
\end{eqnarray}
and all physical states must satisfy the constraint of gauge
invariance
\begin{equation}
G^a(x)\Psi[A]=\left[\partial_iE^a_i(x)-gf^{abc}A^b_i(x)E^c_i(x)\right]
\Psi[A]=0
\label{constr}
\end{equation}
Under a gauge transformation $U$ (generated by $G^a(x)$) the
vector potential transforms as
\begin{equation}
A^a_i(x)\rightarrow  \ A^{Ua}_i(x)=S^{ab}(x)A_i^b(x)+\lambda_i^a(x)
\label{gt}
\end{equation}
where
\begin{eqnarray}
S^{ab}(x)={1\over 2}tr\left(\tau^aU^\dagger\tau^bU\right);
\ \ \ \lambda_i^a(x)={i\over g}tr\left(\tau^aU^\dagger
\partial_iU\right)
\label{defin}
\end{eqnarray}
and $\tau^a$ are traceless Hermitian N by N matrices satisfying
$tr(\tau^a\tau^b)=2\delta^{ab}$. A  gaussian wave functional
\begin{eqnarray}
\Psi[A_i^a]=exp\left\{ - \frac{1}{2}\int d^{3}x d^{3}y
 \left[A_i^a(x)-\zeta_i^a(x)\right]
 (G^{-1})^{ab}_{ij}(x,y)
 \left[A_j^b(y)-\zeta_j^b(y)\right]\right\}
\end{eqnarray}
transforms under the gauge transformation as
\begin{equation}
\Psi[A_i^a]\rightarrow\Psi[(A^{U})_i^{a}]
\end{equation}
In the abelian case it is enough to take $\partial_iG^{-1}_{ij}=0$
to satisfy the constraint of gauge invariance,  In the nonabelian
case, however, due to the homogeneous piece in the gauge transformation
(\ref{gt}), no gauge invariant Gaussian WF exist.

One possible strategy is to disregard this fact \cite{kerman} and
hope that one does not loose much by minimizing the energy in the
whole Hilbert space, which also includes unphysical states. This
is, however, very risky. The sticking point is that the hamiltonian
of the theory is unique {\it only} on physical states. One can
add to equation (\ref{ham}) an arbitrary operator multiplied by one
of the generators of the gauge group without changing the
energies of the physical states, but reshuffling the rest of
the spectrum beyond recognition. In this way the gap between
the physical vacuum and some of the unphysical states can be
made very small. In fact the large Hilbert space can even
contain states which have energies lower than the physical
vacuum. Since we are working with a particular Hamiltonian, it is
not clear a priori that this is not the case. Therefore
minimizing the energy on the whole space may lead to huge
admixtures of unphysical states in the "best variational state",
making the results of such a procedure meaningless. Of course,
one could be lucky and with the particular choice of the
Hamiltonian (\ref{ham}) all unphysical states may have
large energies, but there is no way to know it without a
separate investigation of this question.

We, at any rate, will restrict our attention to gauge invariant
states only.
It is clear then, that the Gaussian ansatz must be modified.
Several modifications were considered in pevious work. One
obvious possibiity is to restrict classical fields to zero
and insert adjoint Wilson lines in the exponential \cite{greensite},
so that
\begin{equation}
\left[A_i^a(x)-\zeta_i^a(x)\right]
(G^{-1})^{ab}_{ij}(x,y)
\left[A_j^b(y)-\zeta_j^b(y)\right] \rightarrow
B_i^a(x)G^{-1}_{ij}(x-y)B_j^b(y)W^{ab}(C)
\end{equation}
where
$W(C)=P\exp\left(ig\int_C dl_iF^aA_i^a \right)$, and $F^a$
are the generators  of SU(N) in the adjoint representation.
This form, however, makes it practically impossible to perform
explicit calculations, except in the week coupling limit.
Another proposed modification is to multiply the Gaussian by a
finite order polynomial in the fields. In that way gauge invariance
can be maintain to a finite order in the coupling constant
\cite{preparata}. Then, however, it is again not quite clear
to which extent the calculation is nonperturbative.

Instead, we will take a straightforward approach, and simply
project the Gaussian WF onto gauge invariant sector. In this
paper we also restrict ourselves to the case of zero classical
fields ($\zeta=0$).
Our variational ansatz is therefore
\begin{equation}
\Psi[A_i^a]=\int DU(x)
 \exp\left\{-{1\over 2}\int d^{3}x d^{3}y
\ A_i^{Ua}(x)G^{-1ab}_{ij}(x-y)\ A_j^{Ub}(y)\right\}
\label{an}
\end{equation}
with $A_i^{Ua}$ defined in (\ref{gt}) and the integration
is performed over the space of special unitary matrices with
the $SU(N)$ group invariant measure.

Before attempting a calculation with this expression, we will
impose several restrictions on the form of $G$, which will lead
to considerable simplifications. First, we will only consider
matrices $G$ of the form
\begin{equation}
G^{ab}_{ij}(x-y)=\delta^{ab}\delta_{ij}G(x-y)
\label{an1}
\end{equation}
 This form is certainly the right one in the perturbative regime.
In the leading order in perturbation theory, the nonabelian character
of the gauge group is not important, and the integration in
equation (\ref{an}) is basically over the $U(1)^{N^2-1}$ group.
The $\delta^{ab}$ structure is then obvious - there is a
complete democracy between different components of the vector
potential. The $\delta_{ij}$ structure arises in the following
way. If not for the integration over the group, $G^{-1}_{ij}$
would be precisely the (equal time) propagator of the electric
field. However due to the integration over the group, the actual
propagator is the transverse part of $G^{-1}$. It is easy to check
that the longitudinal part $\partial_iG_{ij}^{-1}$ drops out of all
physical quantities. At the perturbative level, therefore, one
can take $G_{ij}\sim \delta_{ij}$ without any loss of generality.
 We will adopt this form of the matrix $G$ also in our variational
calculation.

We can use additional perturbative information to restrict the
form of $G$ even further. The theory of interest is asymptotically
free. This means that the short distance asymptotics of correlation
functions must be the same as in the perturbation theory.
Since $G^{-1}$ in perturbation theory is directly related
to correlation functions of gauge invariant quantities (e.g. $E^2$),
we conclude
\begin{equation}
G^{-1}(x)\rightarrow  {1\over x^4}, ~~~~~ x \rightarrow 0
\label{an2}
\end{equation}

Finally, we expect the theory nonperturbatively to have a gap.
In other words, the correlation functions should decay to zero
at some distance scale
\begin{equation}
G(x)\sim 0, \ \ x>{1\over M}
\label{an3}
\end{equation}
We will build this into our variational ansatz in  the
simplest
possible way. We  will take $M$ to be our only variational
parameter.
This can be done by choosing for $G(x)$ a particular form
that has
the UV and IR asymptotics
given by  (\ref{an2}) and  (\ref{an3}), like, for example a massive
scalar propagator with mass $M$. We find another parametrization
slightly more convenient. The form that will be used throughout
this calculation has the following Fourier transform:
\begin{eqnarray}
G^{-1}(k) = \left\{ \begin{array}{ll} \sqrt{ k^{2} ~} &
\mbox{ if  $ k^2>M^2$}\\
 M &  \mbox{ if $k^2<M^2$}
\end{array}
\right.
\label{an4}
\end{eqnarray}
We have checked, that using a massive propagator instead,
practically does not change the results.
Equation (\ref{an})  together with equations (\ref{an1}) and (\ref{an4})
  define our  variational ansatz. We now have to calculate
  the energy expectation  value in these states and minimize it with
 respect to the only variational parameter left - the scale $M$.
 Note, that the perturbative  vacuum is included in this set of states,
and corresponds to $M=0$.
A nonzero result for $M$ would therefore mean a nonperturbative
dynamical scale generation in the Yang - Mills vacuum.
In the next section we will explain the approximation scheme we use
to calculate expectation values in the trial state.

\newsection{The effective sigma model and the renormalization group.}

The question now is, how do we calculate expectation values in the
state of the form (\ref{an}):
\begin{eqnarray}
<O>&=& {1\over Z}\int DU DU' <O>_{A}\\
 <O>_{A}&=& \int DA e^{-{1\over 2}
\int dx dy  A_i^{Ua}(x)G^{-1}(x-y)A_i^{Ua}(y)}~ O~
e^{-{1\over 2}\int dx' dy' A_j^{U'b}(x')G^{-1}(x'-y')A_j^{U'b}(y')}\nonumber
\end{eqnarray}
where $Z$ is the norm of the trial state. Two simplifications are
immediatelly obvious. First, since we will only be considering gauge
invariant operators $O$, one of the group integrations is redundant.
Performing the change of variables $A\rightarrow A^U$ (and remembering
that both inegration measures
$DU$ and $DA$ are group invariant), we obtain (omitting the volume
 of $SU(N)$ factor $\int dU$)
\begin{eqnarray}
<O>&=& {1\over Z}\int DU <O>_{A}\\
<O>_{A}&=& \int DA e^{-{1\over 2}\int dx dy A_i^{Ua}(x)
G^{-1}(x-y) A_i^{Ua}(y) }O e^{-{1\over 2}\int dx' dy'
A_j^b(x')G^{-1}(x'-y') A_j^b(y')}\nonumber
\nonumber
\end{eqnarray}
Also, since the gauge transform of a vector potential is a linear
function of $A$ (\ref{gt}), for fixed $U(x)$ this is a Gaussian
integral, and can therefore be performed explicitly for any reasonable
operator $O$. We are left then only with a path integral over one
group variable $U(x)$. But this is a tough one!

Let us consider first the normalization factor $Z$. After
integrating over the vector potential we obtain:
\begin{equation}
Z=\int DU\exp\{-\Gamma[U]\}
\label{sigma}
\end{equation}
with an action
\begin{equation}
\Gamma[U]={1\over 2} Tr \ln{\cal M}
+{1\over 2}\lambda[G+SGS^T]^{-1}\lambda
\label{action}
\end{equation}
where multiplication is understood as the matrix
multiplication with indices: colour $a$, space $i$ and position
(the values of space  coordinates) $x$,
i.e.
\begin{equation}
(A B)_{ik}^{ac}(x,z) = \int d^{3}y A_{ij}^{ab}(x,y)
 B_{jk}^{bc}(y,z),~~~~~~~~~~
 \lambda O \lambda = \int d^3x d^3y \lambda^{a}_{i}(x) O^{ij}_{ab}(x-y)
 \lambda^{b}_{j}(y)
\end{equation}
 The trace $ Tr $ is understood as a trace over  all three types of indices.
  In equation (\ref{action})  we have defined
\begin{equation}
S^{ab}_{ij}(x,y)=S^{ab}(x)\delta_{ij}\delta(x-y), \ \
{\cal  M}^{ab}_{ij}(x,y)=
[S^{Tac}(x)S^{cb}(y)+\delta^{ab}]G^{-1}(x-y)\delta_{ij}
\label{def1}
\end{equation}
where $
S^{ab}(x)={1\over 2}tr\left(\tau^aU^\dagger\tau^bU\right)$ and
$~\lambda_i^a(x)={i\over g}tr\left(\tau^aU^\dagger
\partial_iU\right)
$
 were defined in (\ref{defin}) and $tr$ is a trace over colour
 indices only.  Using the completeness condition for $SU(N)$
\begin{equation}
\tau^{a}_{ij}\tau^{a}_{kl} = 2 \left(\delta_{il}\delta_{jk} -
 \frac{1}{N}\delta_{ij}\delta_{kl}\right)
\label{completeness}
\end{equation}
one can see that $S^{ab}$ is an orthogonal matrix
\begin{equation}
S^{ab}S^{cb} = \frac{1}{4} \tau^{b}_{ij}\tau^{b}_{kl}
\left(U\tau^aU^\dagger\right)_{ji}
\left(U\tau^cU^\dagger\right)_{kl} = \frac{1}{2}tr\left(\tau^{a}
\tau^{b}\right) = \delta^{ab}
\label{orthogonal}
\end{equation}
where we used that $tr\left(U\tau^cU^\dagger\right) =
tr \tau^c =  0$.

We have written action (\ref{action}) in a form which suggests a
convenient
way of thinking about the problem. The path integral (\ref{sigma})
 defines a
partition function of a nonlinear sigma model with the target
space $SU(N)/Z_N$ in three dimesnional Euclidean space. The fact
that the target space is $SU(N)/Z_N$ rather than $SU(N)$, follows
from the observation that the action (\ref{action})  is invariant
under local transformations belonging to the center of $SU(N)$.
This can be trivially traced back to invariance of $A_i^a$ under
gauge transformations that belong to the center of the gauge
group.

We note, that the quantity $U(x)$ has a well defined gauge
invariant meaning, and it is {\it not} itself a matrix of
a gauge transformation. A contribution of a given $U(x)$ to
the partition function eq.\ref{sigma} and to other expectation
values corresponds to the contirbution to the same quantity
from the off diagonal matrix element between the initial
Gaussian and the Gaussian gauge rotated by $U(x)$. Therefore,
if matrices $U(x)$ which are far from unity give significant
contribution to the partition function, it means that the off
diagonal contribution is large, and therefore that the
simpleminded non gauge invariant Gaussian approximation
(which neglects the off diagonal elements) misses important
physics.

The action of this sigma model is rather complicated.
 It is a nonlocal and a nonpolynomial functional of $U(x)$.
There are however two observations, that will help us devise an
approximation scheme to deal with the problem. First, remembering
that the bare coupling constant of the Yang Mills theory is small,
let us see how does it enter the sigma model action. It is easy to
see, that the only place it enters is the second term in
 the action (\ref{action}), because $\lambda^{a}_{i}(x)$ has an explicit
 factor $1/g$.
 Moreover, it enters in the same way as a
coupling constant in a standard sigma model action. We can
therefore easily set up a perturbation theory in our sigma
model.
With the standard parametrization
\begin{equation}
U(x)=\exp\{i{g\over 2}\phi^a\tau^a\}
\end{equation}
one gets $\lambda^{a}_{i}(x) = - \partial_{i}\phi^a(x) + O(g),~~~
 S^{ab}(x) = \delta^{ab} + O(g)$ and
the leading order term in the action becomes:
\begin{equation}
{1\over 16}\int d^{3}xd^{3}y
\partial_i\phi^a(x)G^{-1}(x-y)\partial_i\phi^a(y)
\end{equation}
This is just a free theory, except that the propagator is
nonstandard, and at large momenta its Fourier transform
behaves like
\begin{equation}
D(k)\sim G(k) {1\over k^{2}} \sim {1\over |k|^3}
\label{prop}
\end{equation}
Nevertheless, the perturbation theory is straightforward. Indeed,
it is easy to see, that in this sigma model perturbation theory
the coupling constant renormalizes logarithmically. The first
order diagram that contributes to the coupling constant
renormalization is the tadpole. In a sigma model
with a standard kinetic term this diagram diverges linearly
 as $\int d^{3}k/k^{2}$,
a sign of perturbative nonrenormalizability. In our model,
though, due to a nonstandard form of the kinetic term
(\ref{prop}), the diagram diverges only logarithmically as
  $\int d^{3}k/k^{3}$.
The form of the $\beta$ function therefore is very similar
to the $\beta$ function in the ordinary QCD perturbation
theory. In this paper we assume, that to one loop, the two
$\beta$ functions indeed coinside.
The explicit calculation in the framework of the sigma model
will be presented  elsewhere \cite{inprog}.
The perturbation theory, therefore becomes worse and worse
as we go to lower momenta, and at some point becomes
inapplicable.

Now, however, let's look at the other side of the coin.
Let us see how does the action look like for the matrices
$U(x)$, which are slowly varying in space. Due to the short
rangedness of $G(x)$, clearly for $U(x)$ which contain only
momenta lower than the variational scale $M$ the action is
local.
In fact, with our ansatz (\ref{an4}) it becomes the
standard action
\begin{equation}
\Gamma_L[U]
={M \over 2g^2} tr \int d^{3}x ~ \partial_iU^\dagger(x)\partial_iU(x)
+ \ldots
\end{equation}
where we omit the higher order in $g$ terms. We have used
the completeness condition (\ref{completeness}) and the
fact that $tr(U^\dagger\partial_iU) = 0$  to rewrite
\begin{equation}
\lambda^{a}_{i}(x)\lambda^{a}_{i}(x) = -(1/g^2)
tr\left(\tau^aU^\dagger\partial_iU\right)
tr\left(\tau^aU^\dagger\partial_iU\right)
= -(2/g^{2}) tr\left(U^\dagger\partial_iU \
U^\dagger\partial_iU\right)
\end{equation}
 In this low-momentum approximation we  also neglected  the
 space dependence of $S^{ab}_{ij}(x)$ in the term $SGS^T$ in
 (\ref{action}), then using the fact that $S$ is an orthogonal
 matrix (\ref{orthogonal}) one gets $SGS^T \rightarrow G$.

Strictly speaking, due to the $Z_N$ local symmetry of the
original theory (\ref{action}), the action for the low
momentum modes is slightly different. The derivatives should
be understood as $Z_N$ covariant derivatives. The most convenient
way to write this action, would be to understand $U(x)$ as
belonging to $U(N)$ rather than $SU(N)$ and introduce a $U(1)$
gauge field by:
\begin{equation}
\Gamma_L={1\over 2}{M \over g^2} tr
 \int d^{3}x ~(\partial_i-iA_i)U^\dagger(x)(\partial_i+iA_i)U(x)
\label{low}
\end{equation}
This defines
a sigma model on the target space $U(N)/U(1)$,
which is isomorfic to $SU(N)/Z_N$.
This action does not look too bad. Even though it still can not
be solved exactly, it is amenable to analysis by standard methods,
such as the mean field approximation, which in 3 dimensions and for
large number of fields should give reliable results.

The suggestion therefore, is the following. Let us integrate
perturbatively the high momentum modes of the field $U(x)$.
This is the renormalization group (RG) transformation. We would
like to integrate out all modes with momenta $k^2>M^2$. This
procedure will necessarilly generate a {\it local} effective
action for the low momentum modes. At the same time,
 because of the (presumable) equivalence of the RG flows in QCD
 and our effective sigma-model,  the effective coupling constant
 will be the the  running QCD coupling constant $\alpha_{QCD}(M)$
 at scale $M$. This part of the theory can
then be solved in the mean field approximation.
Clearly, in order for the perturbative RG transformation to be
justified, the QCD running coupling constant at the scale $M$
must be small enough. Our procedure will then make sense, provided
the energy will be minimized at the value of the variational
parameter, for which
\begin{equation}
\alpha_{QCD}(M)<1
\label{consistency}
\end{equation}
We will check whether this consistency condition is satisfied at
the end of the calculation.
In the next section we will calculate the expectation value of
the Hamiltonian in the lowest order of this approximation scheme,
and perform the minimization with respect to $M$.

Before doing that, we would like to make one side remark. It is
amusing to see how  the present framework can accomodate
instanton effects. Recall, that in a path integral formalism
 instantons describe the tunneling transition between  some initial
 state $\Phi[A]$ and a new state $\Phi[\tilde{A}]$ where a field
 $\tilde{A}_{i}$ is obtained from a field  $A_{i}$ by a
 large gauge  transformation which is described by a nontrivial element
  of the  homotopy group $\Pi_3(SU(N)/Z_N)=Z$.
  The target space of the effective sigma model $SU(N)/Z_N$, has the right
topology:  $\Pi_3(SU(N)/Z_N)=Z$. The model therefore
must have classical "hedgehog" solutions analogous to Skyrmions
\cite{skyrme}.
In fact in the perturbative regime they should be easy to find.
At week coupling the action reduces to (up to a numerical coefficient)
$\int d^{3}x d^{3}y
 tr \left[U^{\dagger}(x)\partial_{i}U(x){1\over (x-y)^4}
  U^{\dagger}(y)\partial_{i}U(y)\right]$,
and equation of motion for $U(x)$ becomes relatively  simple.
Note also, that
this action has a dilatational invariance $x \rightarrow \lambda x$,
Skyrmion solution must approach $1$ asymptotically at large
distances.  These functions $U_{cl}(x)$ then correspond to contributions
of the off diagonal matrix elements between the initial Gaussian,
and the same Gaussian
gauge transformed by a large gauge transformation, which is precisely
the meaning of one instanton contribution to the path integral.

Note that the dilatational invariance is broken in our ansatz for
slowly varying modes, by apearance of the scale $M$. Indeed, the
only Skyrmion solutions in the low momentum effective action
(\ref{low}),  are pointlike, due to Derrick's collapse. This means
physically, that the scale $M$ sets the nonperturbative infrared
limit on the instanton size. Our variational vacuum therefore, is
free from the infrared problem associated with the large size
instantons.

The  variational ansatz which has been considered corresponds to a
 zero value of the QCD $\theta$ - parameter, since we have integrated
over the entire gauge group without any extra phases. As
is well known, the general $\theta$-vacuum is defined as
\begin{equation}
|\theta> = \sum_{n} e^{i n \theta} |n>
\label{theta}
\end{equation}
where $n$ labels the topological sectors in the configuraion
 space (space of all potentials $A_{i}^{a}(x)$).  Generalization
 of our trial wave functions
to nonzero $\theta$ is trivial - all we need to do
 is to insert in equation (\ref{an})
 an extra  phase factor in the integrand
\begin{equation}
\exp\left\{i{\theta\over 24\pi^2}\int dx \epsilon_{ijk}
tr\left[(U^\dagger\partial_iU)
(U^\dagger\partial_jU)(U^\dagger\partial_kU)\right]\right\}
\label{thetafactor}
\end{equation}
The integrand here is a properly normalized topological charge,
and
it takes integer values for topologicaly nontrivial configurations
$U(x)$, i.e. this factor reproduces the $\exp(in\theta)$ term in
 (\ref{theta}). This phase factor can be obtained also if one
 remembers that usually the $\theta$-dependence of the wave functional
 is given by the $\exp\left[i\theta S_{CS}(A) \right]$, where
 $ S_{CS}(A)$ is a Chern-Simons term, which under the gauge
 transformation $U$ transforms as
\begin{equation}
S_{CS}(A^{U}) = S_{CS}(A) +
{1\over 24\pi^2}\int dx \epsilon_{ijk}
tr\left[(U^\dagger\partial_iU)
(U^\dagger\partial_jU)(U^\dagger\partial_kU)\right]
\end{equation}
thus integrating over $U$ leads precisely
to the phase factor (\ref{thetafactor}).
The state thus constructed, is an eigenstate of an operator of the
large gauge transformation with eigenvalue $e^{i\theta}$.
This will result in addition of the same topological term to
the effective action (\ref{action}). It is amusing that for
$\theta=\pi$, the "Skyrmions" in the effective theory  will
be "fermions".
In the rest of this paper, we shall ignore instanton contributions,
 but it will be interesting to come back to this question later.

\newsection{Solving the variational equation.}

We will now calculate the expectation value of the energy.
\begin{eqnarray}
H&=&{1\over 2}\int d^{3}x E_i^{a2}+{1\over 2} \int d^{3}x
(\epsilon_{ijk}
\partial_jA^a_k)^2+\nonumber \\
&+&{1\over 2}g \epsilon_{ijk}\epsilon_{ilm}f^{abc}
\int d^{3}x \partial_jA_k^aA_l^bA_m^c+
{g^2\over 8}\epsilon_{ijk}\epsilon_{ilm}f^{abc}f^{ade}
\int d^{3}x  A_j^bA_k^cA_l^dA_m^e
\end{eqnarray}

We first perform the Gaussian integrals over the vector potential
at fixed $U(x)$. Let us consider, for example, the calculation of
 the chromoelectric energy:
\begin{eqnarray}
\int d^{3}x< E_i^{a2}>_{A} = \int d^{3}x
< - \frac{\delta}{\delta A_{i}^{a}(x)}
\frac{\delta}{\delta A_{i}^{a}(x)}>_{A} = \nonumber \\
 Tr G^{-1} - \int d^{3}x d^{3}y d^{3}z
 G^{-1}(x-y)  G^{-1}(x-z) < A_{i}^{a}(y) A_{i}^{a}(z)>_{A}
\end{eqnarray}
  Using (\ref{action}) it is easy to calculate the
 average over $A$. Defining for convenience
\begin{equation}
a_i^a(x)=
\int d^{3}yd^{3}z
\lambda_i^b(y)G^{-1}(y-z)S^{bc}(z)({\cal M}^{-1})^{ca}(z,x)
\label{a}
\end{equation}
so that  gaussian integration over $A$ is $\int DA
\exp\left[ - (1/2)(A+a){\cal M} (A+a)\right]$ one
 gets
\begin{eqnarray}
\int d^{3}x< E_{i}^{a~2}>_{A}  =
3(N^2-1)\int d^{3}x  G^{-1}(x,x)-
\int d^{3}x (G^{-1}{\cal M}^{-1}G^{-1})^{aa}_{ii}
(x,x) \\
- \int d^{3}x d^{3}y  a_i^a(x)G^{-2}(x-y)a_i^a(y) \nonumber
\end{eqnarray}
where $G^{-2}(x-y) = \int d^{3}z G^{-1}(x-z)G^{-1}(z-y)$ and
 ${\cal M}^{-1}$ is defined as $\int d^{3}y {\cal M}^{-1}(x,y)
{\cal M}(y,z) = \delta^{3}(x-z)$. Let us note that $G^{-2}$
 has dimension $[x]^{-5}$ and ${\cal M}^{-1}$ has dimensiion
 $[x]^{-2}$.
 For chromomagnetic field  the calculations  are
 straightforward  and one gets
\begin{eqnarray}
<(\epsilon_{ijk}\partial_jA^a_k)^2>_{A}  =
(\epsilon_{ijk} \partial_ja^a_k)^2
+
\epsilon_{ijk}\epsilon_{ilm}
 \partial^x_i\partial^y_l
({\cal M}^{-1})^{aa}_{km}(x,y)|_{x=y} ~~~~~~
\end{eqnarray}
\begin{eqnarray}
 <\partial_jA_k^aA_l^bA_m^c>_{A} &  = &
\partial_ja_k^aa_l^ba_m^c+
\partial_ja^a_k({\cal M}^{-1})^{bc}_{lm}(x,x) \\
& + &  a^b_l\partial_j^x({\cal M}^{-1})^{ac}_{km}(x,y)|_{x=y}+
a^c_m\partial_j^x({\cal M}^{-1})^{ab}_{kl}(x,y)|_{x=y}
\nonumber
\end{eqnarray}
\begin{eqnarray}
\epsilon_{ijk}\epsilon_{ilm}f^{abc}f^{ade}<A_j^bA_k^c
A_l^dA_m^e>_{A} =
2f^{abc}f^{ade}a_j^ba_k^ca_l^da_m^e ~~~~~~~~~~~~~~~~~~ \\
+
8f^{abc}f^{ade}a_i^ba_i^d ({\cal M}^{-1})^{ce}(x,x)
+12f^{abc}f^{ade}({\cal M}^{-1})^{bd}(x,x)
({\cal M}^{-1})^{ce}(x,x)
 \nonumber
\end{eqnarray}
Here we have used the obvious notation
${\cal M}^{ab}_{ij}={\cal M}^{ab}\delta_{ij}$
The next step is to decompose the matrix field $U(x)$
into low and high momentum modes. In general this is a
nontrivial problem. However, since we are only going to
integrate over the high momenta in the lowest order in
perturbation theory, for the purposes of our
calculation we can write
\begin{equation}
U(x)=U_L(x)U_H(x)
\end{equation}
where $U_L$ contains only modes with momenta $k^2<M^2$, and
$U_H$ has the form $U_H=1+ig\tau^a\phi_H^a$ and $\phi_H$
contains only momenta $k^2>M^2$.
This decomposition is convenient, since it preserves the group
structure. Also, since the measure $DU$ is group invariant,
we can write it as $DU_LDU_H$. With this decomposition
we have:
\begin{equation}
\lambda_i^a(x)=S_H^{ab}(x)\lambda_{iL}^b(x)+\lambda_{iH}^a(x)
\end{equation}
Further simplifications arise, since we only have to keep the
leading piece in $\phi^a_H$. We can therefore write in our
approximation:
\begin{eqnarray}
S^{ab}(x)&=&S^{ab}_L(x)\nonumber \\
{\cal M}^{ab}(x,y)&=&2\delta^{ab}G^{-1}(x-y)\nonumber \\
\lambda_i^a(x)&=&\lambda_{iL}^a(x)+\lambda_{iH}^a(x)\nonumber \\
a^a_i(x)&=&{1\over 2}\lambda_{iL}^a(x)+{1\over 2}
\lambda_{iH}^b(x)S_L^{ba}(x)
\label{simple}
\end{eqnarray}
We are now in the position to rewrite different pieces in the
VEV of energy in this approximation:
\begin{eqnarray}
\int d^{3}x < E_{i}^{a~2}>_{A}={3(N^2 - 1)\over 2}\int G^{-1}(x,x)
{}~~~~~~~~~~~~~~~ \nonumber \\ - {1\over 4}
\int d^{3}xd^{3}y\lambda_{iL}^a(x)G^{-2}(x-y)\lambda_{iL}^a(y)
- {1\over 4}\int d^{3}xd^{3}y
\lambda_{iH}^a(x)G^{-2}(x-y)\lambda_{iH}^a(y)
\label{esq}
\end{eqnarray}
The cross term vanishes, since to this order, as we shall
see , there is a decoupling between the high and the low
momentum modes in the action, and therefore the product
factorizes, and $<\lambda_{iH}^a>=0$.
Our ansatz for $G^{-1}$ (\ref{an4}) allows us to simplify
this expression further. Remember that $\lambda_L(x)$ contains
only momenta below $M$. Then it is immediate to see that,
\begin{equation}
\int d^{3}xd^{3}y\lambda_{iL}^aG^{-2}(x-y)
\lambda_{iL}^a(y)=M^2\int dx\lambda_{iL}^a(x)\lambda_{iL}^a(x)
\end{equation}
We can then rewrite equation  (\ref{esq}) as
\begin{eqnarray}
\int d^{3}x < E_{i}^{a~2}>_{A}  = {3(N^2-1)\over 2}
\int G^{-1}(x,x)-  \nonumber \\
{M^{2}\over 4}\int d^{3}x
\lambda_{iL}^a(x)\lambda_{iL}^a(x)
 - {1\over 4}\int d^{3}xd^{3}y
\lambda_{iH}^a(x)G^{-2}(x-y)\lambda_{iH}^a(y)
\label{esq1}
\end{eqnarray}
The contribution of the magnetic term to the energy is very simple.
All cross terms between the low and high momentum modes drop out.
Some vanish for the same reason as the cross terms in equation
 (\ref{esq}),
and others because they are explicitly multiplied
by a power of the coupling constant. Since our approximation
is the lowest
order in $g$, except for the nonanalytic contributions that
come from the low mode effective action, those terms do not
contribute. In fact, the entire low momentum mode contribution
drops out of this term. The reason is that the only terms which
could give a leading order contribution, is
\begin{equation}
\int (\epsilon_{ijk}\partial_j\lambda^a_{kL})^2
\end{equation}
It can be rewritten as
\begin{equation}
(f^{a}_{ijL})^2+O(g^2)
\end{equation}
Where $f^{a}_{ijL}$ is the "magnetic field" corresponding to
"vector potential" $\lambda^a_{iL}$.
However, $\lambda_L$ has the form of a pure gauge vector potential.
Therefore
$f^{a}_{ijL}=0$, and the contribution of this term is higher order
in $g^2$. We have checked, that including this term, indeed changes
the energy density in the best variational state by a small amount
($O(10\%)$), but has no effect at all on the best value of the
variational parameter $M$.
The entire magnetic field contribution to the energy is then:
\begin{equation}
{1\over 2}< B^2>_{A}={1\over 8}
 (\epsilon_{ijk}\partial_j
\lambda^a_{kH})^2+{N^2-1\over 2} \partial_i^x\partial_i^yG(x-y)|_{x=y}
\end{equation}
The last step is to perform an averaging over the $U$-field.
For convenience, we rewrite here the complete expression for
the energy density (here $V = \int d^{3}x$ is a space volume):
\begin{eqnarray}
{<2H>\over V} &=&{3(N^2-1)\over 2} G^{-1}(x,x) +
(N^2-1) \partial_i^x\partial_i^yG(x-y)|_{x=y} \nonumber \\
&-&{1\over 4 V}\int d^{3}xd^{3}y<\lambda_{iH}^a(x)G^{-2}(x-y)
\lambda_{iH}^a(y)>_U
+{1\over 4} <(\epsilon_{ijk}\partial_j \lambda^a_{kH})^2>_U
\nonumber \\
&-&{M^{2}\over 4 V}\int d^{3}x
<\lambda_{iL}^a(x)\lambda_{iL}^a(x)>_U
\label{energy}
\end{eqnarray}
 where  the averaging  over the $U$ - field
should be performed with the
sigma model action (\ref{action}).
In our approximation this action has a simple form.
Using equation  (\ref{simple})  we obtain
\begin{equation}
\Gamma ={1\over 4}\int dx dy \lambda^a_{iH}(x)G^{-1}(x-y)
\lambda^a_{iH}(y)+
{M\over 4}\int dx \lambda^a_{iL}(x)\lambda^a_{iL}(x)
\label{hl}
\end{equation}
The low momentum mode part is precisely equal to $\Gamma_L$ in
equation (\ref{low}).
The only difference is, that the coupling constant that appears
in this action should be understood as the running coupling
constant at the scale $M$. This, obviously is the only $O(0)$
effect of the high momentum modes on the low momentum
effective action.
\begin{equation}
\Gamma_L=
{1\over 2}{M\over g^2(M)}
 tr \int d^{3}x (\partial_i-iA_i)U^\dagger(x)(\partial_i+iA_i)U(x)
\label{low1}
\end{equation}

We are now in a position to evaluate the VEV of energy.
The contribution of the high momentum modes is immediatelly
calculable. Using the parametrization $U_H(x)=1-{i\over 2}g\phi^a\tau^a$,
we find that $\phi^a$ are free fields with the propagator
\begin{equation}
<\phi^a(x)\phi^b(y)>=2\delta^{ab}[\partial^x_i\partial^y_i
G^{-1}(x-y)]^{-1}|_{p^2>M^2}
\label{phiphi}
\end{equation}
Also to this order
$\lambda^a_{iH}(x)=\partial_i\phi^a(x)$ and
 therefore $\epsilon_{ijk}\partial_j\lambda^a_{kH}=0$.
Using (\ref{phiphi}) one can see that
\begin{equation}
{1\over 4 }\int d^{3}xd^{3}y<\lambda_{iH}^a(x)G^{-2}(x-y)
\lambda_{iH}^a(y)>_U = V {N^2 -1 \over 2} \int_{M}^{\Lambda}
{d^3k\over (2\pi)^3} G^{-1}(k)
\end{equation}
 where  $\Lambda$ is the ultraviolet cutoff,
  and the the contribution of the high momentum modes to the energy
(first  two lines in equation (\ref{energy})) is:
\begin{eqnarray}
{2E_0 \over V}&=&
(N^2-1)\left\{\int_0^\Lambda {d^3k\over (2\pi)^3}\left[G^{-1}(k)+
k^2G(k)\right]+{1\over 2}\int_0^M{d^3k\over (2\pi)^3}G^{-1}(k)
\right\}\nonumber \\
&=&  {N^2-1\over 2\pi^2}\left\{
\int_{0}^{M} k^{2} dk \left[{3\over 2} M + {k^{2} \over M}\right]
 + 2\int_{M}^{\Lambda} k^{3} dk  \right\} =
{N^2-1\over 10\pi^2}M^4+ ...
\label{ehigh}
\end{eqnarray}
 Terms denoted by $...$ in eq. \ref{ehigh} depend on $\Lambda$,
 but are independent of the variational scale $M$.

We now have to evaluate the contribution of the low momentum
modes. It is clear from the form of the action (\ref{low1}),
that this contribution as a function of $M$ will not be
featureless. The most convenient way to think about it, is
from the point of view of classical statistical mechanics.
Comparing equations (\ref{energy}) and (\ref{low1}), we see that
we have to evaluate the internal energy of the sigma model
(with the UV cutoff $M$) at the temperature proportional to
the running coupling constant $g^2(M)$. For large
$M$\footnote{Large $M$, of course means large relative
to $\Lambda_{QCD}$.}, the coupling constant is small,
which corresponds to the low temperature regime of the
sigma model. In this regime the global $SU(N)\otimes SU(N)$
symmetry group of the model is spontaneously broken. Lowering
$M$, we raise $g^2(M)$, and therefore the temperature. At some
critical value $g_C$, the model will undergo a phase transition
into the unbroken (disordered) phase. Clearly, in the vicinity
of the phase transition all thermodynamical quantities will
vary rapidly, and therefore this is a potentially interesting
region of coupling constants.

Before analysing the phase transition region let us calculate
$E(M)$ for large $M$. In this regime the low momentum theory is
weekly coupled. The calculation is straightforward, and to
lowest order in $g^2$ gives:
\begin{equation}
{1\over 4}M^2<\lambda^a_{iL}(x)\lambda^a_{iL}(x)>=
{N^2-1\over 12\pi^2}M^4
\label{elow}
\end{equation}
Putting this together with the high momentum contribution, we find
\begin{equation}
{E(M)\over V}= {N^2-1\over 120\pi^2}M^4, ~~~~~~~
 M>>\Lambda_{QCD}
\label{largeM}
\end{equation}
This indeed is the expected result. The energy density monotonically
increases as $M^4$, with the slope which is given by the standard
perturbative expression. Note, however that the slope is very
small, and the contribution of the low momentum modes to the
energy is negative. Therefore, if the internal energy of the
sigma model grows significantly in the phase transition region,
the sign of $E(M)$ could be reversed\footnote{The energy, of
course never becomes negative, since equation (\ref{ehigh})
 contains  a divergent $M$ - independent piece. Here we consentrate only
on the $M$-dependence of $E$.}  and the energy will then be
minimized for $M$ in this region.

To see, whether this indeed happens, we will now study the low
momentum sigma model in the mean field approximation.
We rewrite the partition function by introducing a (hermitian
matrix) auxiliary field
$\sigma$ which imposes a unitarity constraint on $U(x)$
\begin{eqnarray}
Z=\int DU D\sigma DA_i \exp\left(- \Gamma[U,A,\sigma]\right)
{}~~~~~~~~~~~~~~~~~~~~~~~~~~~
\\
\Gamma[U,A,\sigma] = {M\over 2g^2(M)}
 tr \int d^{3}x \left[\left(\partial_i-iA_i\right)
U^\dagger(x)\left(\partial_i+iA_i\right)U(x) + \sigma
\left(U^\dagger U-1\right)\right] \nonumber
\end{eqnarray}
The role of the vector field $A_i$ is to impose a $U(1)$ gauge
invariance, and thereby to eliminate one degree of freedom.
As far as the thermodynamical properties are concerned, its
effect is only felt as an $O(1/N^2)$ correction. At the level
of accuracy of the mean field approximation, we can safely
disregard it, which we do in the following.
The mean field equations are:
\begin{equation}
<U^\dagger U>=1
\label{one}
\end{equation}
\begin{equation}
<\sigma U>=0
\label{two}
\end{equation}
 From equation (\ref{two}) it follows that either $<\sigma>=0$,
$<U>\ne 0$ (the ordered, broken symmetry phase with massless
Goldstone bosons),  or $<\sigma>\ne0$, $<U>= 0$
(the disordered, unbroken phase with massive excitations).
We are mostly interested in the disordered phase, since there the mean
field approximation should be reliable. Since the symmetry is
unbroken, the expectation value of $\sigma$ should be proportional to
 a unit matrix
\begin{equation}
<\sigma_{\alpha\beta}>=\sigma^2 1_{\alpha\beta}
\end{equation}
Equation (\ref{one}) then becomes
\begin{equation}
2N^2{g^2(M)\over M}\int_0^M{d^3k\over (2\pi)^3}{1\over k^2+\sigma^2}=
{N^{2} g^2(M)\over \pi^2} \left(1 - {\sigma\over M}
\arctan{M\over \sigma}\right) = N
\label{gap}
\end{equation}
The gap equation (\ref{gap}) has solution only for couplings
 (temperatures)  $g^{2}(M)$
 larger than the
 critical coupling (temperature) $g^{2}_{C}$, which
is determined by the
condition that $\sigma = 0$
\begin{equation}
\alpha_C={g^2_C\over 4\pi}={\pi \over 4}{1\over N}
\label{gc}
\end{equation}
The low momentum mode contribution to the ground state energy is
\begin{equation}
N^2M\int_0^M{d^3k\over (2\pi)^3}{k^2\over k^2+\sigma^2}=
{N^2\over 2\pi^2}M\left[{1\over 3}M^3-\sigma^2M+\sigma^3
\arctan{M\over \sigma}\right]
\end{equation}

The final mean field expression for the ground state energy
density
is (we do not distinguish between $N^{2}$ and $N^{2} - 1$ since
we have neglected the contribution of the $U(1)$ gauge field - the
errors are of order
$1/N^{2}$ and are definitely smaller than the error introduced by
using the mean field approximation in the first place)
\begin{equation}
E= {N^2\over 4\pi^2}M^4\left[-{2\over 15}+{\sigma^2\over M^2}
{\alpha_C\over \alpha(M)}\right]
\label{fen}
\end{equation}
where $\alpha(M)$ is the QCD coupling at the scale $M$, $\alpha_C$
is given by equation (\ref{gc}), and $\sigma$ is determined by
\begin{equation}
{\sigma\over M}\arctan {M\over \sigma}={\alpha(M)-\alpha_C\over \alpha(M)}
\end{equation}

The energy as a function of $M$  is plotted on Fig.1 for $N=3$.
Qualitatively
it is the same for any $N$.
The minimum of the energy is obviously at the point
$\alpha(M)=\alpha_C$.
Using the one loop Yang - Mills $\beta$ function and
$\Lambda_{QCD}=150 Mev$, we find for $N=3$
\begin{equation}
M=\Lambda_{QCD}e^{24\over 11}=8.86\Lambda_{QCD}=1.33 {\rm Gev}
\end{equation}

To see, what is the phenomenological significance of this number
we have calculated the value of the gluon condensate
 $(\alpha/\pi)<F_{\mu\nu}^{a}F_{\mu\nu}^{a}> =
 (2\alpha / \pi)\left(< B^{a~2}_{i}> - <E^{a~2}_{i}>\right)$.
After some  calculations which
 are  straightforward and do not contain any new ingredients,
 we get\footnote{We have again kept only the $M$ - dependent pieces.
Each one of the quantities $< E^{a~2}_{i}>$ and $< B^{a~2}_{i}>$ is
of course positive, due to positive UV divergent, but $M$ independent
pieces. It is easy to check that the energy density
 $ E = 1/2(< E^{a~2}_{i}> + < B^{a~2}_{i}>) = - (1/30\pi^{2})
 N^{2} M^4$ coincides with the first term in equation
 (\ref{fen}), as it must.}
\begin{equation}
< E^{a~2}_{i}> = -{1 \over 24 \pi^2} N^{2} M^4, ~~~~~~~
< B^{a~2}_{i}> = -{1 \over 40 \pi^2} N^{2} M^4
\end{equation}
 and finally  obtain:
\begin{equation}
{\alpha\over \pi}<F_{\mu\nu}^{a}F_{\mu\nu}^{a}>=
 N M^{4} {1\over 2}\left(-{1 \over 40 \pi^2} +
{1 \over 24 \pi^2}\right) = {N\over 120\pi^2}M^4=0.008~ Gev^4
\end{equation}

The best phenomenological value of this condensate, is
$0.012$ Gev$^4$ \cite {sumrules}.
Considering, that $<F^2>$ is proportional to the fourth power of
$M$, our result is very reasonable. For example, changing $M$ by
only $10\%$,
from $1.33$ Gev to $1.46$ Gev would give
$(\alpha/ \pi)<F^2>=0.0116$  Gev$^4$, in perfect
agreement
with \cite{sumrules}.

Note that for $N=3$, the value of the $QCD$ coupling constant
at the variational scale is $\alpha_C=0.26$. It is reasonably small,
so that the consistency condition for the perturbative integration
of the high momentum modes is satisfied. However, it is not so small
that higher order corrections
be negligible. We expect therefore that including higher orders in
perturbation theory can give corrections to our result for $\alpha(M)$
of order $25\%$.
Since $M$ depends exponentially on $\alpha(M)$, such change in
$\alpha$ may change the value of $M$ by a factor of $2-3$.
Consequently our result for $F^2$ should be taken only as an
order of magnitude estimate. This is usually the case in
theories with logarithmically running coupling constants.
 The best accuracy is always achieved for dimensionless
quantities, since those usually
are slowly varying functions of $\alpha$. The overall scale
depends on $\alpha$ exponentially, and therefore always has the
largest error. It would be interesting to calculate some
dimensionless quantities, such as the ratio of the square of
the string tension to the SVZ condensate, in our approach \cite{inprog}.

Another uncertainty comes from the use of the mean field
approximation.
As a rule, mean field approximation gives a good estimate of the
critical temperature. Sometimes, however it gives wrong predictions
for the oder of the phase transition. We believe that this is indeed
the case here. Our results would indicate that the phase transition
is second order. The mass gap in the sigma model vanishes continuously
at the critical point. The universality class describing the symmetry
breaking pattern $\left(SU(N)\otimes SU(N)\right)/ SU(N)$ was
considered in
the context of finite temperature chiral phase transition in QCD.
The results of $\epsilon$ - expansion \cite{pw} and also numerical
simulations \cite{phtl} strongly suggest that the phase transition
is of  first order. In our case there is an additional $Z_N$ symmetry in
the game. However, if anything, we believe that its presence should
increase the latent heat rather than turn the transition into a
second order one. The reason is, that the $Z_N$ gauge invariant
theory allows existence of topological defects - the $Z_N$ strings,
and condensation of topological defects frequently leads to
discontinuous phase transitions.

Nevertheless we believe that our results are robust against this
uncertainty.
The mean field approximation should be reliable in the regime where
the mass gap in the sigma model is not too small. At the point $M=4.5
\Lambda_{QCD}$ we find
\begin{equation}
\sigma=0.23 M,  \ \ \ \ \alpha(M)=0.38
\end{equation}
Since the gap is of the order of the UV cutoff, the mean field
approximation is reliable in the vicinity of this point. The
perturbation theory is also still reasonable at this value of $\alpha$.
The fact, that the energy is negative and has a
minimum for some $\alpha(M)<.38$, seems to be
therefore unambiguous.

We now want to argue, that independently of the mean field calculation,
it is physically very plausible that the energy is minimized precisely
at the critical temperature, on the disordered side of the phase
transition (if it is of the  first order).
Consider first, the contribution of the high momentum modes to the
ground state energy, equation (\ref{ehigh}). It is proportional to $M^4$
with a fixed ($M$ - independent) proportionality coefficient
$x= (N^2-1)/ 10\pi^2 $.  Consider now the low momentum contribution
in the large $M$ region equation (\ref{elow}).
 It is again proportional to
$M^4$ with the coefficient $y_0=(N^2-1) 12\pi^2$. The
proportionality coefficient of the low momentum contribution at
the phase transition point, according to our calculation is twice
as big $y_C= 2N^2/12\pi^2 $ (we disregard the difference
between $N^2$ and $N^2-1$).
This is very easy to understand physically. In the large $M$ - low
temperature regime the global symmetry of the sigma model
$SU(N)\otimes SU(N)$ is broken down spontaneously to $SU(N)$.
This leads to appearance of $N^2-1$ massless Goldstone bosons.
In fact, at zero temperature, those are the only propagating degrees
of freedom in the model. All the rest have masses of the order of
UV cutoff, and therefore do not give any contribution to the
internal energy.
Now, when the temperature is raised ($M$ is lowered), the
Goldstone bosons remain massless and other excitations become
lighter. If the transition is second order, at the phase
transition point the symmetry is restored, and one should
have  a complete multiplet of the $SU(N)\otimes SU(N)$ symmetry
of massless particles. The dimensionality of this multiplet is
$2(N^2-1)$. The contribution of every degree of freedom to the
internal energy is still roughly the same as at zero temperature.
This is so, since, although at the phase transition the particles
are interacting, critical exponents of scalar theories in 3 dimensions
are generally very close to their values in a free theory \cite{blgzj}.
The internal energy at this point therefore should be roughly twice
its value at zero temperature.
Moving now to higher temperatures, all the particles become heavier,
and therefore their contribution to internal energy decreases. The
internal energy therefore should have a maximum at the phase
transition temperature.

Note, that the ground state energy of the Yang - Mills theory, is
the difference between the high, momentum contributions and the
internal energy of the low mode sigma model. Already at zero
temperature, these two contributions differ only by $20\%$, and
that is why the coefficient in the
expression equation (\ref{largeM}), even though positive, is so small.
At the critical point, where the low momentum mode internal energy
is twice as large, the chances of the slope becoming negative are
very good. This is indeed precisely what happens in our mean field
analysis, but according to the previous argument this in large
measure is independent of the approximation.
If the phase transition is first order one should be more careful.
The internal energy then changes discontinuously across the phase
transition.
The particles in the disordered phase are always massive, and the
internal energy is smaller that in the case of the second order
phase transition. However, if the transition is only weakly first
order the same argument
still goes through (the fact that the mean field predicts second
order phase transition, may be an indication that if it is in fact
first order, it is only weakly so). In fact, it does seem very likely
that the ground state energy will become negative, since all is need
for that, is that the sigma model internal energy grows by $20\%$ at
the phase transition relative to the zero temperature limit. Moreover,
in this case there will be a finite latent heat, which means that the
internal energy in the disordered (high temperature) phase is higher.
The ground state energy, therefore, will have its minimum in the
disordered phase.

We believe, therefore, that our results are qualitatively correct,
and will survive the improvement of the approximation.

\newsection{Wilson loop and area law}

 The next interesting question is, whether the
variational state we found describes the physics of confinement.
The relevant quantity to calculate is the Wilson loop
\begin{equation}
W(C) = <tr~ P \exp\left(i{g\over 2}\oint_C dx_i A_{i}^{a} \tau^{a}
\right)>
\label{wloop}
\end{equation}
When averaging over $A$ we must take into account the $P$-ordering
 of the exponent - the simplest way to do it is to introduce
   new degrees of freedom living on the contour $C$ which, after
 quantization, become the $SU(N)$ matrices $\tau^{a}$
 \cite{wilsonloop}. We shall consider here how it works in the
 case of $SU(2)$ group -  the generalization of this construction to
an arbitrary Lie group has been discussed in \cite{wilsonloop}.

 The construction is based on the observation made in
 \cite{pas} that instead of considering the ordered product
 of $\tau^{a}$ matrices one can consider the correlation function
\begin{eqnarray}
<{\tau^{a}(t_{1})\over 2} {\tau^{b}(t_{2})\over 2}
\ldots {\tau^{c}(t_{k})\over 2}> \longrightarrow
<n^{a}(t_{1}) n^{b}(t_{2})\ldots n^{c}(t_{k})> =
{}~~~~~~~~~~~~~~ \\
\int Dn(t) n^{a}(t_{1}) n^{b}(t_{2})\ldots n^{c}(t_{k})
\exp\left[i(S+1/2) \int_{\Sigma} d^{2}\xi
\epsilon_{\mu\nu}\epsilon^{abc} n^{a}
\partial_{\mu}n^{b}\partial_{\nu}n^{c}\right] \nonumber
\end{eqnarray}
where $S$ is the spin of representaion, i.e. for fundamental
 representation $S = 1/2$, $n^{a}(t)$ is a unit vector
 $n^{a}n^{a} = 1$ living on a contour $C$ ($t$ is a coordinate
 on the contour) and $\Sigma$ is an arbitrary two-dimensional
 surface with the boundary $C = \delta\Sigma$. The
 two-dimensional action (here and later
 we shall concentrate only on case $S=1/2$)
\begin{equation}
S[n] = \int_{\Sigma} d^{2}\xi
\epsilon_{\mu\nu}\epsilon^{abc} n^{a}
\partial_{\mu}n^{b}\partial_{\nu}n^{c}
\label{S[n]}
\end{equation}
 depends only on values $n^{a}(t)$
 at the boundary. The variation of the action is
\begin{equation}
\delta S = \oint_{C} dt \epsilon^{abc} n^{a}
\partial_{t}n^{b} \delta n^{c}
\label{Svariation}
\end{equation}
Here we have used the fact that $\delta n^{a}n^a=0$
(because $n^{c}n^{c} = 1$) and thus
 $\epsilon_{\mu\nu}\epsilon^{abc}
\partial_{\mu}n^{a}\partial_{\nu} n^{b}\delta n^{c} = 0$.
 It is easy to
 see that $<n^{a}(t_{1}) n^{b}(t_{2})\ldots n^{c}(t_{k})>$
 depends only on the ordering of $t_{1},\ldots t_{k}$ - as it
 should. To see  this and the fact that $n^{a}(t)$ behaves
 effectively as $\tau^{a}$ let us make   local   field
 reparametrization
\begin{equation}
n^{a}(t) \rightarrow n^{a}(t) + \epsilon^{abc} \Omega^{b}(t)
n^{c}(t)
\end{equation}
under which the action variation (\ref{Svariation})
  is $\delta S = -
\oint_{C} dt \dot{n}^{a}(t)\Omega^{a}(t)$ and one gets
 the Ward identities (it is important to remember here, that
correlators in
 any QFT are averages of the $T$-ordered products)
\begin{eqnarray}
\frac{d}{dt}<n^{a}(t) n^{b}(t_{1})\ldots n^{c}(t_{k})>= i
\sum_{i=1}^{k}\delta(t-t_{i})\epsilon^{adf}<n^{f}(t)
n^{b}(t_{1}) \ldots \underline{n^{d}(t_{i})}
\ldots  n^{c}(t_{k})>
\label{ward}
\end{eqnarray}
where $\underline{n^{d}(t_{i})}$ means the exclusion of this term
 from the products of the fields in a correlator. From
(\ref{ward}) one can conclude immediately that correlation
function indeed  depends only on ordering of $t_{1},\ldots t_{k}$
and the following equal time commutaion relations hold
\begin{equation}
\left[n^{a}, n^{b}\right] = i \epsilon^{abc} n^{c}
\end{equation}
 which means that one can substitute $n^{a}$ by  a Pauli
 matrix $\tau^{a}/2$.
As a result one can represent the Wilson loop (\ref{wloop}) in
the form
\begin{equation}
W(C) = <\int Dn(t)
\exp\left[i\int_{\Sigma} d^{2}\xi
\epsilon_{\mu\nu}\epsilon^{abc} n^{a}
\partial_{\mu}n^{b}\partial_{\nu}n^{c}\right]
 \exp\left(
ig\oint_C dx_i A_{i}^{a}(x(t)) n^{a}(t)
\right)>
\end{equation}
and now we can average over $A_{i}$ using (\ref{action})
 and (\ref{a})
\begin{eqnarray}
<\exp\left(
ig\oint_C dx_i A_{i}^{a}(x(t)) n^{a}(t)
\right)>_{A} = \exp\left(-
ig\oint_C dx_i a_{i}^{a}(x(t)) n^{a}(t)
\right) \nonumber \\
 \exp\left(-{1\over2}\oint_C\oint_Cdt_{1}dt_{2}\dot{x}_{i}(t_1)
\dot{y}_{i}(t_2)n^{a}(t_1)n^{b}(t_2)({\cal M}^{-1})^{ab}(x,y)
\right)
\end{eqnarray}
where $a_{i}^{a}$  was defined in (\ref{a}).
Now the Wilson loop can be calculated as the average over
 two scalar fields: $U(x)$ living in the whole space and
 $n^{a}(\xi)$ living on a two-dimensional surface $\Sigma$
 such that $C = \delta \Sigma$
\begin{eqnarray}
W(C) = \int DU \int Dn \exp\left(-\Gamma[U] + iS[n]\right)
\exp\left(-
ig\oint_C dx_i a_{i}^{a}(x(t)) n^{a}(t)
\right) \nonumber \\
 \exp\left(-{1\over2}\oint_C\oint_Cdt_{1}dt_{2}\dot{x}_{i}(t_1)
\dot{y}_{i}(t_2)n^{a}(t_1)n^{b}(t_2)({\cal M}^{-1})^{ab}(x,y)
\right)
\label{wloopnu}
\end{eqnarray}
In the infrared limit, which is of main
 interest to us here,
 we can use (\ref{simple}) to
 simplify (\ref{wloopnu}) and get
\begin{eqnarray}
W(C) = \int DU_{L} \int Dn \exp\left(-\Gamma_{L}[U] + iS[n]\right)
\exp\left(-
i{g\over 2}\oint_C dx_i \lambda_{i~L}^{a}(x(t)) n^{a}(t)
\right) \nonumber \\
 \exp\left(-{1\over4}\oint_C\oint_Cdt_{1}dt_{2}\dot{x}_{i}(t_1)
\dot{y}_{i}(t_2)n^{a}(t_1)n^{a}(t_2)G(x-y)
\right) \\
\int DU_{H}\exp\left(-\Gamma_{H}[U]\right)
\exp\left(-
i{g\over 2}\oint_C dx_i \lambda_{i~H}^{b}(x(t))S^{ba}_{L} n^{a}(t)
\right) \nonumber
\label{wloopnu1}
\end{eqnarray}
Using equation  (\ref{phiphi}) one can see that the last term in
  in (\ref{wloopnu1}) after integrating
 over the $U_{H}$  becomes equal to the second term
 and one gets finally
\begin{eqnarray}
W(C) =
\int Dn \exp\left(iS[n]\right)
 \exp\left(-{1\over2}\oint_C\oint_Cdt_{1}dt_{2}\dot{x}_{i}(t_1)
\dot{y}_{i}(t_2)n^{a}(t_1)n^{a}(t_2)G(x-y)
\right)
 \nonumber \\
\int DU \exp\left(-\Gamma[U]\right)
\exp\left[
{1\over 2}\oint_C dx_i tr\left(\tau^{a}U^\dagger\partial_iU\right)
 n^{a}(t)\right] ~~~~~~~~~~~~~~~~~
\label{wloopfinal}
\end{eqnarray}
where integrating $DU$ is over  low-energy modes only and
$\Gamma[U]$ is the corresponding low-energy action.
Since $G(x-y)$ is short range, the term
\begin{equation}
\exp\left(-{1\over2}\oint_C\oint_Cdt_{1}dt_{2}\dot{x}_{i}(t_1)
\dot{y}_{i}(t_2)n^{a}(t_1)n^{a}(t_2)G(x-y)
\right)
\end{equation}
gives only perimeter dependence and one can neglect it
when calculating the string tension. Then it can be shown,
rewriting $n^{a}(t)$ as
 the $\tau^{a}$  and performing some  simple algebra, that  the
calculation of the Wilson loop is closely related
 to the calculation of  the  vacuum expectation value of the
monodromy operator
\begin{equation}
M=tr P\exp\left(\oint_C dl_iU^\dagger\partial_iU\right)
\end{equation}
in the low momentum sigma
model with an effective action $\Gamma[U]$\footnote{In fact,
the Wilson loop does not reduce to $M$, but rather to $tr
P\exp\left(\frac{1}{2}\oint_C dl_iU^\dagger\partial_iU\right)$.
We believe, however, that qualitatively its behaviour should
be similar.}
Since the target space of the sigma model is ${\cal M}=SU(N)/Z_N$,
and $\Pi_1({\cal M})=Z_N$, this factor can take on values
$\exp{i2\pi n/N}$.
It has a natural interpretation in terms of the topological
defects in the sigma model. As mentioned already, the topology
allows existence of $Z_N$ strings. The string creation operator
and the operator $M$ satisfy the commutation relations of the
t'Hooft algebra \cite{tHooft}. Therefore, in the presence of a
string, the operator $M$ has expectation value $\exp{i2\pi n/N}$,
where $n$ is the linking number between the loop $C$ and the string.
As we have argued, the sigma model is in the disordered phase. Usually,
this means that the topological defects are condensed. The vacuum of
the sigma model must have therefore a large number of strings, and
the VEV of $\cal M$, probably, will average to zero very quickly,
and for large loops will have an area law
$W(C) \sim \exp\left(-\alpha' A\right)$.
Stricktly speaking, for  this to happen, one needs not only a large
number of strings, but also a large fluctuations in this number,
but those, usually come  together.

We also would like to mention that the model of two
fields - $U$ and $n^{a}$ - defined in (\ref{wloopfinal}) is of
some interest in itself. For example one can study how nonperturbative
fluctuations of both fields - $Z_{N}$ strings and
Skyrmions for $U$ and instantons
for $n$ interact with each other - these questions as well as
a calculation of $\alpha'$  will be
considered in \cite{inprog}.

An interesting point is, that if one couples fundamental fermions to
the Yang Mills fields, the effective sigma model will not have a $Z_N$
gauge symmetry any more. The origin of this $Z_N$ symmetry is the
fact, that the Yang Mills fields do not transform under the center
of the gauge group. Fundamental fermions, however, do transform
nontrivially, and therefore the sigma model action will depend on
these matrices $U$. The target space now therefore is $SU(N)$,
rather than $SU(N)/Z_N$, and is simply connected. The topology of
the target space does not allow strings any more. Therefore, if
it is true, that it is the condensation of these objects, that
is responsible for the area law for the Wilson loop, the area law
will disappear. This is in complete agreement with one's
expectations, that in a theory with fundamental charges, an
external test charge can be screened, and therefore there is
no area law for the Wilson loop.

\newsection{Discussion and  Conclusion}

In this paper we have presented a simple variational calculation of
the Yang Mills ground state WF. Our trial states preserved gauge
invariance explicitly.
The results are encouraging. We find that the energy is minimal for
a state which is different from the perturbative vacuum, even though
the perturbative vacuum state was included in our variational ansatz.
Dynamical scale generation takes place and the gluon (SVZ)
 condensate in the  best variational state is nonzero.

It is interesting to note, that from the point of view of the
effective sigma model, the energy is minimized in the disordered
(unbroken) phase. In other words, the fluctuations of the field $U$
are big, unlike in the perturbative regime (high momentum modes),
where $U$ is very close to a unit matrix.
 From the point of view of the original WF this means that the off
diagonal contributions, coming from the Gaussian gauge rotated by
a slowly varying gauge transformations, are large. This is telling
us, that it was indeed necessary to project the initial Gaussian
onto gauge invariant state. Without doing this, we would miss the
contributions of the off diagonal elements to the energy
expectation value.

There is still a lot of work to be done, even in the framework of
our variational ansatz. Our present paper should be considered only as
an
exploratory research.
Of course, coupling the fermions is a very interesting question in
itself. It seems to us, that it should be possible to treat a theory
with fermions in basically the same variational approach as presented
here. It would be then very interesting to see the chiral symmetry
breaking and calculate fermionic condensates.

Quite apart from this, there are several technical points that can
be improved. First, we are planning to extend the RG calculation to
take into account the one loop contribution of the high momentum
modes. This might require to change a variational ansatz slightly.
One may have to consider not the gauge projected Gaussians, but
gauge projected products of Gaussians and polynomials of the fourth
order in the fields. This does not change the level of complexity of
the calculation.

It would also be desirable to have better methods to deal with the
low momentum sigma model, especially since we suspect that the mean
field approximation does not give the correct order of the phase
transition.
Although we do not expect the variational parameter to be very
sensitive to this, the vacuum condensates can depend strongly on
the mass gap of the sigma model.

Finally, there is one more direction, in which the calculation can
be extended. In this paper we have adopted the simplest ansatz for
the width of the Gaussian $G$, based on the argument, that it should
be short ranged. The Fourier transform of our propagator goes to a
constant at zero momentum. This, however is no the only possible
form of a short range correlator. It could have a different small
momentum behaviour. It is quite possible that the small momentum
behaviour is very important. One could therefore introduce an
additional variational parameter $\gamma$, assuming the asymptotic
small momentum dependence of the function $G$ to be of the form
$k^\gamma$. This will only affect the last step of our calculation.
The action of the effective low momentum model will have extra
derivatives.

In conclusion, it seems to us that the type of the variational
approximation presented here is manageable, and also gives some
preliminary interesting results. It therefore warrants further
work along the lines described in this section.

{\bf Acknowledgements.}
I.I.K. thanks the members of T-8
theory group at LANL for warm hospitality
during his visit in February 1994, when  the first
impulse for this work has been done.
A.K. is grateful to the
members of T-8, and especially to
Tanmoy Bhattacharya, Fred Cooper, Yuval Kluger and Michael Mattis,
for many interesing conversations and valuable comments and for
their patience during the weeks he was constantly bugging them
in the T-8 corridor.
We thank Bill Bardeen, David Gross, Arthur Kerman, Alex Krasnitz, Eduardo
Marino, Sasha Migdal, Sasha  Polyakov, Baruch
Rosenstein, Misha  Shifman and Arkadyi Vainshtein for interesting
discussions
The work of I.I.K. was supported by  NSF grant No. PHY90-21984.

\noindent

\bigskip

{\renewcommand{\Large}{\normalsize}

\newpage
Figure Caption.

Fig.1  The energy density of a variational state as a function of
the variational parameter $M$ in units of $\Lambda_{QCD}$.
The energy is only shown for $M<8.86\Lambda_{QCD}$, which corresponds
to the disordered phase of the effective low momentum $\sigma$ model.
Close to the phase transition point in the ordered phase, the mean
field approximation is not applicable. Far from the phase transition
point, at large $M$ the energy density is a monotonically increasing
function of $M$ given in equation (\ref{largeM}).


\end{document}